%% file: Sustainability_considerations.tex
\documentclass[11pt]{article}
\usepackage{graphicx}
\usepackage[margin=1.25in]{geometry}
\usepackage[usenames,dvipsnames]{color}
\usepackage{url}
\usepackage[colorlinks = true,
            linkcolor = blue,
            urlcolor  = blue,
            citecolor = blue,
            anchorcolor = blue]{hyperref}

\newcommand\pubnumber{Preprint number }
\newcommand\pubdate{\today}

\def\Title#1{\begin{center} {\LARGE #1 } \end{center}}
\def\Author#1{\begin{center}{ \sc #1} \end{center}}
\def\Address#1{\begin{center}{ \it #1} \end{center}}

\newcommand\pubblock{\rightline{\begin{tabular}{l} \pubnumber\\
         \pubdate \end{tabular}}}
\newenvironment{Abstract}{\begin{quotation} \begin{center}
                       ABSTRACT
     \end{center}\bigskip  }{\end{quotation}}

\input workshopsymbols.tex

\newcommand\snowmass{\begin{center}\rule[-0.2in]{\hsize}{0.01in}\\\rule{\hsize}{0.01in}\\
\vskip 0.1in Submitted to the  Proceedings of the US Community Study\\ 
on the Future of Particle Physics (Snowmass 2021)\\ 
\rule{\hsize}{0.01in}\\\rule[+0.2in]{\hsize}{0.01in} \end{center}}

\begin{document}

\pubblock

\Title{Sustainability Considerations for Accelerator and Collider Facilities}

\bigskip 

\Author{Thomas Roser$^a$ on behalf of the ICFA Panel for Sustainable Accelerators and Colliders}

\medskip

\Address{$^a$BNL, Upton, New York, USA}

\medskip

 \begin{Abstract}
\noindent As the next generation of large accelerator-based facilities are being considered at the Snowmass 2021 study high priority has to be given to environmental sustainability including energy consumption, natural resource use and the environmental impact of effluents. Typically, increased performance – higher beam energies and intensities – of proposed new facilities have come with  increased electric power consumption. In the following we discuss the most important areas of development for the sustainability of accelerator-based research infrastructures in three categories - technologies, concepts and general aspects. To achieve the goal of increased performance with reduced energy consumption a focused R\&D effort is required with the same or even higher priority as the traditional performance-related R\&D.  Such a recommendation was included in the recent European Strategy for Particle Physics Accelerator R\&D Roadmap \cite{European}.
\end{Abstract}

\snowmass

\def\thefootnote{\fnsymbol{footnote}}
\setcounter{footnote}{0}

\section{Introduction}

Scarcity of resources, along with climate change originating from the excessive exploitation of fossil energy are ever growing concerns for humankind. Particularly, the total electric power consumption of scientific facility operations will become more important as the reliance on fossil fuels is being reduced, carbon-neutral energy sources are still being developed and a larger part of the energy consumption, in particular transportation, is converted from fossil fuel to electric power. 

In our accelerator community we need to give high priority to the realization of sustainable concepts, particularly when the next generation of large accelerator-based facilities is considered. Indeed, the much-increased performance – higher beam energy and intensity – of proposed new facilities comes together with anticipated increased electric power consumption. In the following we classify the most important development areas for sustainability of accelerator driven research infrastructures in three categories - technologies, concepts and general aspects. We suggest investing R\&D efforts in these areas and to assess energy efficiency with an equal level of relevance as the classical performance parameters of the facilities under discussion.

\section{Energy efficient technologies}

Energy efficient technologies have a long history in the accelerator facilities for particle physics since often the required performance could only be reached with highly energy efficient devices such as superconducting magnets and superconducting RF cavities. Below are some items, where R\&D could further improve energy efficiency. 

{\it Low loss superconducting resonators:} Cryogenic losses in superconducting resonators can be significant for linacs, particularly in CW operation. The R\&D on high Q superconducting resonators should be continued with high priority. Resonators using Nb3Sn-coating have shown good performance  \cite{Liepe} and could be operated at 4.5 K. At this temperature the cryogenic efficiency is much improved, while still reasonable Q values are achieved. 

{\it Efficient Radio Frequency (RF) sources:} For many accelerators the main power flow involves converting grid power to RF power. To improve the overall efficiency RF sources must be optimized. Efforts should be invested for efficient klystron concepts (e.g. adiabatic bunching and superconducting coils), magnetrons (mode locking) and solid-state amplifiers \cite{EuCARD} \cite{ARIES} \cite{Workshop}. 

{\it Permanent magnets:} Permanent magnets don’t need electrical power. As a side effect no heat is introduced which has a positive effect on the stability of a magnet lattice. Significant progress has been made with permanent magnets for light sources, and for example tunable quadrupoles for the CLIC linacs \cite{EuCARD2}. 

{\it Highly efficient cryogenic systems:} Another important development are efficient cryogenic sys- tems (e.g. He/Ne refrigeration), allowing to optimize heat removal in cold systems from synchrotron radiation and other beam induced energy deposition  \cite{Workshop2}. 

{\it Superconducting electrical links:} Cables using High Temperature Superconductors (HTS) allow to power high-current devices from a distance with no or little losses, thus enabling to install the power converters outside of radiation areas  \cite{Ballarino}. 

{\it Use of heat pumps:} Heat recovery in aquifers is often done at low temperatures with limited usefulness. But after boosting the heat to a higher temperature level using heat pumps, this waste heat can be used for residential heating.

\section{Energy efficient accelerator concepts}

Increasing the energy efficiency of accelerator components can significantly reduce energy consumption, but different accelerator concepts, especially with built-in energy recycling, has the potential to drastically reduce the energy consumption without compromising the performance. 

{\it Energy Recovery Linacs:} The Energy Recovery Linac (ERL) concept was first proposed in 1956 and it allows the recirculation of the beam power after the beam is used by decelerating it in the same RF structures. Using this concept for the electron and positron beam a high energy e+e- collider could be built where more luminosity can be achieved with much less beam intensity than using storage rings since the single beam collision can be much more disruptive. The much lower beam intensity then results in much less energy lost to radiated synchrotron power \cite{Litvinenko}. For a high energy collider, the energy savings can amount to over a 100 MW. In view of the significant technical challenges this scheme should be studied and optimized in more detail. 

{\it Intensity Frontier Machines:} For Intensity Frontier Machines the conversion efficiency of primary beam power for example to Muon/Neutrino beam intensity is a critical parameter. With optimized target and capture schemes the primary beam power, and thus the grid power consumption, can be minimized. Similar arguments are valid for accelerator driven neutron sources  \cite{Workshop}. 

{\it Muon Collider:} For very high parton collision energies the Muon Collider  \cite{Muon} exhibits a favorable scaling of the achievable luminosity per grid power. With constant relative energy spread bunches can be made shorter at higher energies, allowing stronger transverse focusing at the interaction points. Besides other arguments this is an important reason for strengthening R\&D efforts on the muon collider concept. 

{\it Energy Management:} With an increasing fraction of sustainable energy sources like wind and solar power in the future energy mix, the production of energy will fluctuate significantly. One way to mitigate the impact of high energy physic facilities on the public grid is to actively manage their energy consumption using local storage or dynamic operation. Investigation of such concepts should be an integral part of design studies. 

{\it Accelerator Driven Systems (ADS):} Accelerator driven sub-critical reactors can be used to reduce the storage time of radioactive waste (transmutation) of nuclear power stations by orders of magnitude. Such concepts would address an important sustainability problem of nuclear power. The development of high intensity accelerators for ADS has synergies with applications for particle physics or neutron sources. Another innovative accelerator-based transmutation concept using muons is proposed in \cite{Mori}.

\section{General sustainability aspects} 

A carbon footprint analysis in the design phase of a new facility can help to optimize energy consumption for construction and operation. For cooling purposes accelerator facilities typically have significant water consumption. Cooling systems can be optimized to minimize the impact on the environment. For the construction of a facility environment-friendly materials should be identified and used preferentially. The mining of certain materials, in particular rare earths, takes place in some countries under precarious conditions. It is desirable to introduce and comply with certification of the sources of such materials for industrial applications, including the construction of accelerators. A thoughtful life-cycle management of components will minimize waste. Many facilities use helium for cryogenic purposes. Helium is a scarce resource today and with appropriate measures the helium loss in facilities can be minimized. 

Many of these issues are discussed at the workshop series on ’Energy for Sustainable Science at Research Infrastructures’ \cite{Workshop2}.







\end{document}

%% file: workshopsymbols.tex
\def\beq{\begin{equation}}
\def\eeq#1{\label{#1}\end{equation}}
\def\eeqn{\end{equation}}

\newenvironment{Eqnarray}%
   {\arraycolsep 0.14em\begin{eqnarray}}{\end{eqnarray}}
\def\beqa{\begin{Eqnarray}}
\def\eeqa#1{\label{#1}\end{Eqnarray}}
\def\eeqan{\end{Eqnarray}}

\let\bar=\overbar

\def\lsim{\mathrel{\raise.3ex\hbox{$<$\kern-.75em\lower1ex\hbox{$\sim$}}}}
\def\gsim{\mathrel{\raise.3ex\hbox{$>$\kern-.75em\lower1ex\hbox{$\sim$}}}}

\def\del{\partial}
\def\Dslash{\not{\hbox{\kern-4pt $D$}}}
\def\dslash{\not{\hbox{\kern-2pt $\del$}}}
\def\pslash{\not{\hbox{\kern-2pt $p$}}}
\def\ETmiss{\not{\hbox{\kern-4pt $E$}}_T}

\def\Dlr{\mathrel{\raise1.5ex\hbox{$\leftrightarrow$\kern-1em\lower1.5ex\hbox{$D$}}}}

\def\MSB{{\bar{M \kern -2pt S}}}
\def\msb{{\bar{\scriptsize M \kern -1pt S}}}

\def\drb{{\bar{\scriptsize D \kern -1pt R}}}

